\documentclass[%
 reprint,
 amsmath,amssymb,
 aps,
]{revtex4-2}

\usepackage{float}
\usepackage{setspace}

\usepackage{graphicx}
\usepackage{dcolumn}
\usepackage{bm}

\renewcommand{\v}[1]{\ensuremath{\mathbf{#1}}} 
 
\renewcommand\eqref[1]{Eq.\;\ref{#1}} 

\usepackage [english]{babel}
\usepackage [english = american]{csquotes}
\MakeOuterQuote{"}

\usepackage{color}					

\begin{document}

\title{Optimization hardness constrains ecological transients}
\author{William Gilpin}
 \email{wgilpin@utexas.edu}
\affiliation{%
Department of Physics, The University of Texas at Austin, Austin, Texas 78712, USA
}%
\affiliation{%
Oden Institute for Computational Engineering and Sciences, The University of Texas at Austin, Austin, Texas 78712, USA
}%

\date{\today}

\begin{abstract}
Living systems operate far from equilibrium, yet few general frameworks provide global bounds on biological transients. In high-dimensional biological networks like ecosystems, long transients arise from the separate timescales of interactions within versus among subcommunities. Here, we use tools from computational complexity theory to frame equilibration in complex ecosystems as the process of solving an analogue optimization problem. We show that functional redundancies among species in an ecosystem produce difficult, ill-conditioned problems, which physically manifest as transient chaos. We find that the recent success of dimensionality reduction methods in describing ecological dynamics arises due to preconditioning, in which fast relaxation decouples from slow solving timescales. In evolutionary simulations, we show that selection for steady-state species diversity produces ill-conditioning, an effect quantifiable using scaling relations originally derived for numerical analysis of complex optimization problems. Our results demonstrate the physical toll of computational constraints on biological dynamics.
\\\\\\
\textbf{Author Summary:} Distinct species can serve overlapping functions in complex ecosystems. 
For example, multiple cyanobacteria species within a microbial mat might serve to fix nitrogen.
Here, we show mathematically that such functional redundancy can arbitrarily delay an ecosystem’s approach to equilibrium.
We draw a mathematical analogy between this difficult equilibration process, and the complexity of computer algorithms like matrix inversion or numerical optimization.
We show that this computational complexity manifests as a transient chaos in an ecosystem’s dynamics, allowing us to develop scaling laws for the expected length of transients in complex ecosystems.
Transient chaos also produces strong sensitivity on the duration and route that the system takes towards equilibrium, affecting the ecosystem's response to perturbations.
Our results highlight the physical implications of computational complexity for large biological networks.
\end{abstract}

\maketitle

\clearpage

\section{Introduction}

\begin{figure*}
{
\centering
\includegraphics[width=0.7\linewidth]{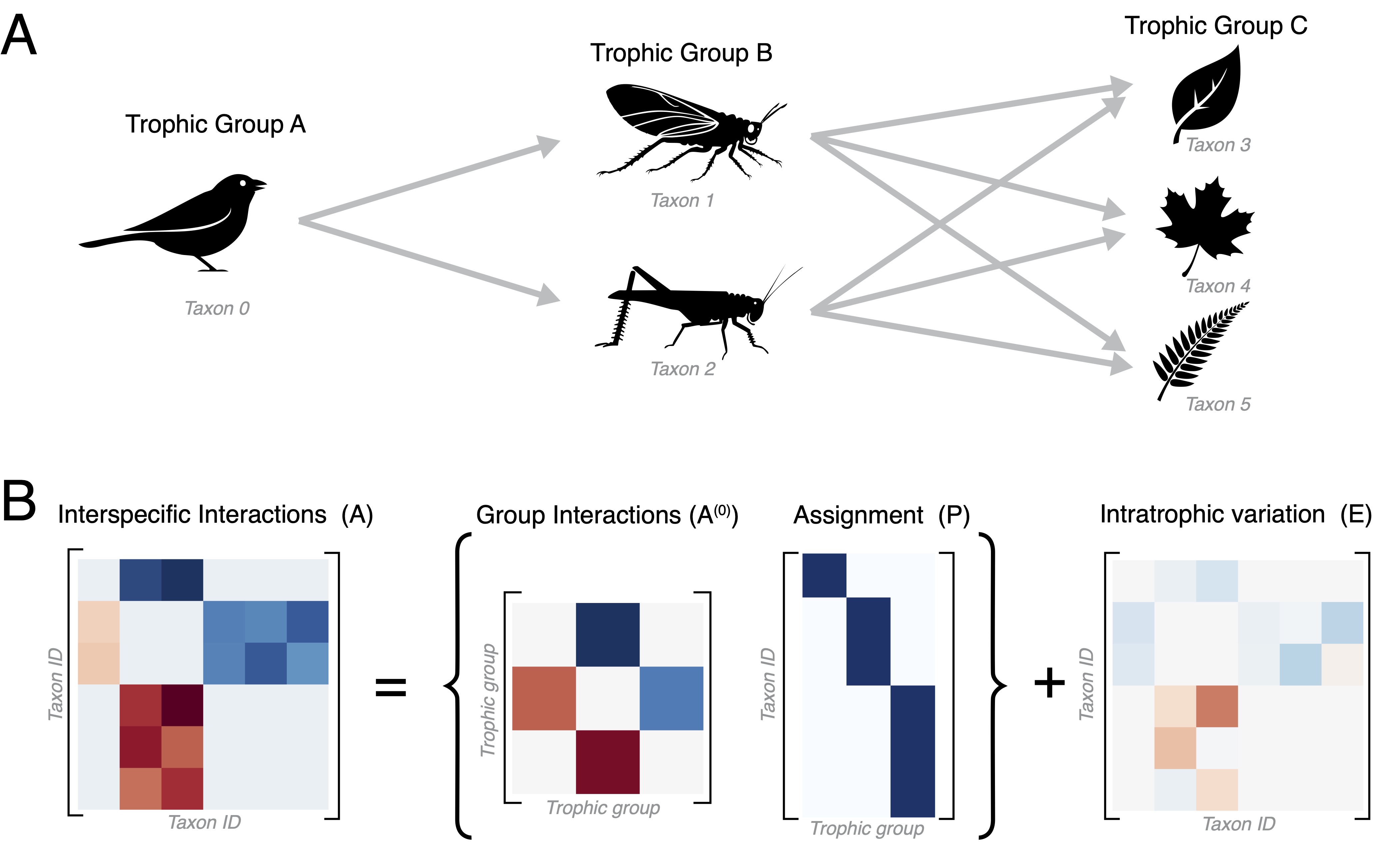}
\caption{
\textbf{Functional redundancy as low-rank interspecific interactions.} 
(A) A three-level food web, in which lower levels contain multiple species that duplicate one other's role. 
(B) The interspecific interaction matrix decomposes into a full-rank group-level component, and a low-rank assignment matrix mapping species to groups. Small-amplitude variations among taxa within each group restore the rank, but lead to high condition number. The exact form of the interaction matrix is given by \eqref{interact}.
}
\label{fig:model}
}
\end{figure*}

Half a century ago, the biologist Robert May used random matrix theory to show that large, random ecosystems are typically unstable---challenging conventional understanding of biodiversity, and suggesting that structural and evolutionary factors tune biological networks towards stable equilibria \cite{may1972will}. But how relevant is stability to real-world biological systems? In physics, high-dimensional dynamical systems typically undergo extended excursions away from stable fixed points. For example, while a fluid flowing through a pipe possesses a globally-stable laminar state, small disturbances trigger transient turbulent excursions with expected lifetimes that increase rapidly with the flow speed \cite{avila2023transition}. 

Transient effects dominate many real-world foodwebs over experimentally-relevant timescales; examples include cyclic succession among rival vegetation species \cite{hastings2018transient}, turnover in patchy phytoplankton communities \cite{morozov2020long}, and establishment of gut microbiota \cite{schlomann2019timescales}. Theoretical works confirm that long-lived transients robustly appear in model foodwebs as the number of interacting species increases \cite{de2024many,ben2021counting,chen2001transient}. Transients may thus influence how real-world ecosystems respond to exogenous perturbations \cite{neubert1997alternatives}, with implications for management and biodiversity \cite{hastings2018transient}. However, analytical techniques primarily consider the effects of small perturbations from equilibrium points, thus characterizing transients in terms of local quantities like reactivity or finite-time Lyapunov exponents \cite{snyder2010makes,townley2007predicting,neubert1997alternatives,tang2014reactivity,suweis2015effect}. 
Such approaches successfully predict experimental phenomena such as critical slowdown near tipping points in microbial ecosystems \cite{dai2012generic,tang2014reactivity}, but they provide less insight into community assembly or large changes (e.g., crises) \cite{hastings2010regime}. Recently, statistical approaches extend local measures by characterizing the \textit{distribution} of fixed points in random ecosystems; these methods calculate the frequency and stability of local minima and marginally-stable equilibria, which act as kinetic traps that impede equilibration \cite{mehta2019constrained,de2024many,tikhonov2017theoretical,biroli2018marginally,blumenthal2024phase}. Statistical approaches also bound the set of valid solutions for a given ecosystem, thus providing null models against which experimentally-observed communities are assessed \cite{tikhonov2017collective,altieri2019constraint,goyal2024universal}. However, few analytical approaches directly characterize the global structure of ecological transients, and it remains unknown whether universal, statistical bounds---analogous to the flow speed in turbulence---govern the onset of transients in ecological networks.

Here, we show that computational complexity measures from optimization theory govern the equilibration of ecological networks. We introduce a class of ill-conditioned ecosystems, for which functional redundancies among species control the rate of approach to equilibrium. We show that the resulting system maps onto a numerical optimization problem, the computational difficulty of which depends on the degree of redundancy. We show that computational complexity leads to transient chaos, in which routes to equilibrium depend sensitively on the initial conditions or assembly sequence of the ecosystem. When dimensionality reduction methods, such as principal components analysis, are applied to the resulting dynamics, they precondition the dynamics by separating fast relaxation from slow "solving" dynamics associated with redundant species. We conclude by using genetic algorithms to evolve random ecosystems to support higher steady-state diversity, which we find increases ill-conditioning and thus optimization hardness. Our results frame ecological dynamics through analogue optimization hardness, and show how computational constraints on biological systems produce physical effects.

\section{Methods}

\begin{figure*}
{
\centering
\includegraphics[width=0.7\linewidth]{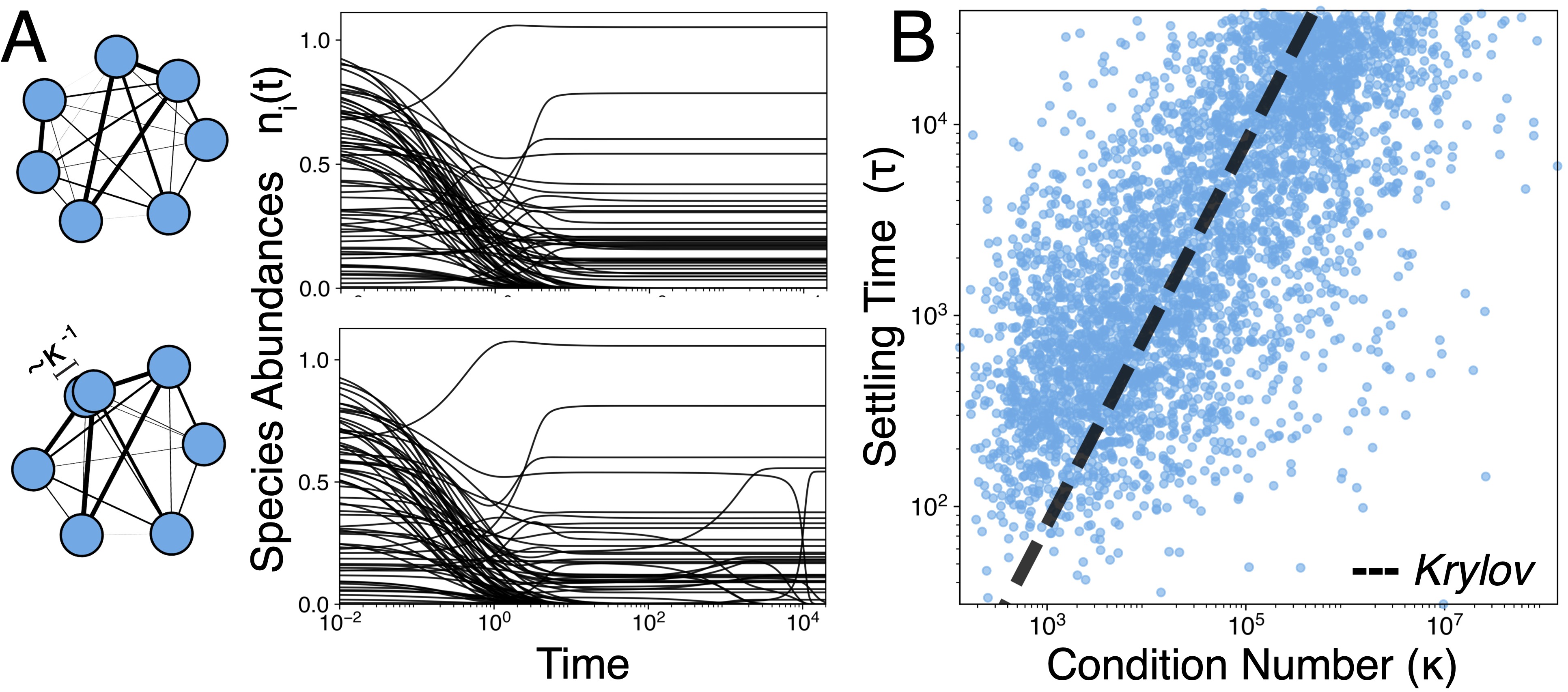}
\caption{
\textbf{Long-lived transients in ill-conditioned ecosystems.}
(A) Equilibration of a random food web without (top) and with (bottom) a pair of functionally-redundant species. Long-lived transients appear in the latter case. (B) Settling time $\tau$ versus condition number $\kappa$ for $10^{4}$ random communities; dashed line shows the scaling expected for an iterative linear program solver.
}
\label{transient}
}
\end{figure*}

We study a minimal model of an ecological network described by the generalized Lotka-Volterra model
\begin{equation}
    \dot{n}_i(t) = n_i(t) \left( r_i + \sum_{j=1}^N A_{ij} n_j(t)  \right),
\label{lv}
\end{equation}
where $n_i(t)$ refers to the abundance of species $i$ with intrinsic growth rate $r_i$. This species interacts with other species in the community through the matrix $A \in \mathbb{R}^{N \times N}$. A full table of symbols and their definitions is given in Appendix \ref{app:symbols}. Each matrix element $A_{ij}$ quantifies the degree to which species $j$ promotes ($A_{ij} > 0$) or inhibits ($A_{ij} < 0$) the growth of species $i$. While \eqref{lv} simplifies the complex mechanics of real-world ecosystems, its general form---growth modulated by pairwise interactions with other species---captures phenomena seen in real ecosystems \cite{servan2018coexistence}.

The interaction parameters $A_{ij}$ are typically unknown for large ecosystems because they require $N(N-1)/2$ separate isolation experiments. Instead, classical works in mathematical biology randomly sample the elements of $A$ from a fixed probability distribution, and then consider the typical behavior of solutions of $A$ averaged across a large sample of possible ecosystems \cite{may1972will}. Typically, $A$ is sampled from the class of matrices,
\begin{equation}
    A = A^{(0)} - d\, I
\label{anull}
\end{equation}
where the elements of $A^{(0)}$ are drawn from a normal distribution, $A^{(0)} \sim \mathcal{N}(0, \sigma)$, $d > 0$ is constant intraspecific density limitation, and $I$ is the identity matrix. The density of a species $i$ governed by \eqref{lv} and \eqref{anull} will initially grow exponentially at a rate $r_i$, before reaching an asymptotic plateau $n_i^*$ at a carrying capacity set by intraspecific crowding or competition ($d$) and a mix of cooperative and competitive interactions with other species ($A^{(0)}$). While \eqref{lv} can, in principle, produce unphysical ecosystems with unbounded exponential growth, when the density limitation $d$ is sufficiently high, \eqref{lv} always has a single stable, non-invasible global equilibrium. This occurs when $A + A^\top$ is negative definite, known as Lyapunov diagonal stability \cite{macarthur1970species}.

The model \eqref{lv} and \eqref{anull} exhibits a known relationship between stability and steady-state diversity as $N$ increases. The global fixed point $n_i(t) \rightarrow n_i^*$ has a quantity $N_\text{coex} \leq N$ species that stably coexist at steady state ($n_i^* > 0$). When \eqref{anull} is resampled many times to produce many replicate ecosystems, the expected number of coexisting species $N_\text{coex}$ obeys a binomial distribution with $\langle N_\text{coex} \rangle_A = N/2$ surviving species on average \cite{servan2018coexistence}. Thus, both steady-state diversity (as measured by $N_\text{coex}$), and variance across different ecosystems, increase with the ecosystem size $N$.

However, prior analyses of \eqref{lv} primarily characterize the equilibrium state $n_i^*$, and not the full dynamics $n_i(t)$. We thus seek to sample a family of interaction matrices that produce dynamical transients. In ecology, transients generically arise due to functional redundancy, in which multiple species have nearly-identical roles in an ecosystem, leading to timescale separation between fast intergroup dynamics and slow intragroup dynamics \cite{louca2016high,morozov2020long,weiner2019spatial}. For example, microbial mats contain multiple similar species performing nitrogen fixation \cite{louca2018function}. These redundant species may slightly differ in morphology or genetic composition, but have identical interactions with all other species \cite{lee2023robustness}. Redundant species may be grouped into metaspecies such as distinct trophic levels, allowing a coarse-grained description of the system in terms of interactions among these groups (Fig. \ref{fig:model}A). However, over long timescales, minute variations lead to slow dynamics within each group. This timescale separation between fast intergroup and slow intragroup dynamics produces long transients.

We introduce redundant structure into \eqref{lv} by sampling $A$ from the family of low-rank matrices,
\begin{equation}
A = P^\top (A^{(0)} - d\, I) P + \epsilon \, E,
\label{interact}
\end{equation} 
where the \textit{assignment matrix} $P \in \mathbb{R}^{N \times N}$ encodes exact functional redundancies, and the \textit{perturbation matrix} $E \in \mathbb{R}^{N \times N}$ encodes small differences among redundant species, which have typical amplitude set by the constant $\epsilon \ll 1$. We draw all parameter values from the unit normal distribution $A^{(0)}_{ij}, E_{ij}, r_i \sim \mathcal{N}(0, 1)$.
When $P = I$, our formulation exactly reduces to prior models \cite{servan2018coexistence}. After sampling $A$, we set $95\%$ of interactions to zero, to mimic the low connectance ($\rho = 0.05$) of real-world communities \cite{dunne2002food}. As in prior works, we find this condition does not affect our observed results as long as the connectance exceeds the percolation threshold $\rho > 1 / N$ (Appendix \ref{app:sparse})\cite{servan2018coexistence}.

When $P \neq I$ and $\text{rank}(P) = M < N$, the rank-deficient assignment matrix $P$ encodes functional redundancy among species in the population. 
For example, in a microbial ecosystem, $N$ corresponds to the full number of resolvable operational taxonomic units (OTU), while $M$ corresponds to the number of functional niches or trophic groups.
In one such scenario considered in previous work, $P$ represents the identity matrix with the $j^{th}$ column duplicated at position $k$, implying that two species $j$ and $k$ serve exactly indistinguishable functions within the ecosystem, such as exactly two nearly-identical nitrogen-fixing species in a microbial food web \cite{tikhonov2017theoretical}. For this particular case, $P$ reduces to a rectangular matrix $P \in \mathbb{R}^{N \times M}$, which maps between $N$ species and $M$ trophic groups. We visualize this case in Figure \ref{fig:model}.

Here, we allow $P$ to encode more general, arbitrary linear combinations of species that collectively substitute for others within the population, such as microbial consortia \cite{louca2016high,taillefumier2017microbial}. When $\epsilon=0$, $P$ and $A$ are rank-deficient and \eqref{lv} is multistable. Each steady-state solution occupies a continuous hyperplane of equivalent solutions containing varying mixtures of the substitutable species. 
For example, if two nitrogen-fixing species are truly indistinguishable, then the dynamics of a microbial food web should be insensitive to one-to-one replacement of one species with the other. 
As a result, across an ensemble of replicate populations, a wide range of functionally-equivalent mixtures of species will be observed.
However, introducing the \textit{singular perturbation} $0 < \epsilon \ll 1$ breaks this symmetry among exact solutions and restores full rank ($\text{rank}(A) = N$), causing the system to regain a single global equilibrium $\mathbf{n}^*$. In the microbial mat example, this corresponds to a minute difference between two nitrogen fixers, leading to gradual competitive dynamics as one species slowly excludes the other. The resulting dynamics exhibit a slow timescale $\sim\! \epsilon^{-1}$ associated with final approach to equilibrium (Fig. \ref{transient}A).

Prior theoretical studies show that steady-state degeneracy due to functional redundancy can be broken by spatial effects, introducing slow dynamical modes associated with a characteristic diffusion time (equivalent to $\epsilon^{-1}$ in our model) \cite{weiner2019spatial}. In the Appendix, we consider another, previously-studied variant of \eqref{lv} that exhibits a transition between multistability and global stability \cite{rieger1989solvable,bunin2017ecological,ros2023generalized}. As the system approaches the transition region, we find that it produces comparable dynamics to our model \eqref{interact}. However, our particular formulation \eqref{interact} maps the onset of complex dynamics onto a single parameter, the redundancy $\epsilon^{-1}$, which we will show has a physical interpretation.

\section{Results}

\subsection{Functional redundancy produces long-lived transients.}

We simulate the dynamics of $10^4$ random ecosystems of the form \eqref{lv}, with interspecific interactions $A$ and functional redundancy $\epsilon^{-1}$ sampled randomly according to \eqref{interact}. Across many random ecosystems, we find that increasing functional redundancy ($\epsilon \rightarrow 0$) delays approach to steady-state, implying that slow competition among similar species limits the rate of equilibration. Varying $\epsilon$ over five orders of magnitude, we observe a power law increase in the settling time $\tau$ that it takes each ecosystem to first reach the global equilibrium $\v{n}^*$ (Fig. \ref{transient}B).

To understand this scaling, we note that the particular form of \eqref{lv} requires that each steady-state solution $\mathbf{n}^*$ solves the constrained linear regression problem $-A \mathbf{n}^* = \mathbf{r}$, $n_i^* \geq 0$. We therefore frame approach to equilibrium in complex ecosystems as a computational challenge: given a set of constraints imposed by the interactions $A$, we seek the corresponding equilibrium $\mathbf{n}^*$ that solves the linear program in \eqref{lv}. 
In the theory of numerical optimization, the intrinsic difficulty of a computational problem determines the minimum time that an algorithm takes to solve it \cite{trefethen2022numerical}.
We thus expect that the lifetime of ecological transients relates to the intrinsic difficulty of the linear optimization problem defined by \eqref{lv}, which is quantified by the condition number, $\kappa(A) \equiv \| A \| \| A^{-1} \|$. A condition number $\kappa \sim 10^0$ implies that $A$ may easily be inverted to solve for $\mathbf{n}^*$. However, as $\kappa \rightarrow \infty$, algorithms for solving the system become increasingly sensitive to small errors; the expected number of digits of required precision increases as $\sim\log_{10}\kappa$ \cite{trefethen2022numerical}. In a continuous natural system like \eqref{lv}, sensitivity at large $\kappa(A)$ leads to transient chaos, which we explore in the next sections.

We expect that ecosystems become ill-conditioned as redundant species become harder to distinguish.  In numerical analysis, iterative solutions of linear programs exhibit a known bound on convergence time $\tau = \log{\xi} / \log[(\kappa - 1) / (\kappa + 1)]$ where $\xi$ is the initial distance from the solution (Fig \ref{transient}B, dashed line) \cite{trefethen2022numerical}. We find that this bound coincides with the scaling of the settling time $\tau$ in random ecosystems, underscoring the physical effects of ill-conditioning in our system. Rather than a purely numerical consideration, $\kappa$ is a physical quantity that can be interpreted as a measure of computational complexity or inverse distance to an ill-conditioned (here, degenerate) problem \cite{demmel1988probability}. As a result, despite the heterogeneity of the various random ecosystems in Fig. \ref{transient}B, the quantity $\kappa$ imposes a global constraint on equilibration time. Thus, while \eqref{lv} represents a model foodweb, we find that analytic predictions can be generated by interpreting ecosystems as computational optimization problems.

\subsection{Dimensionality reduction optimally preconditions the dynamics}

\begin{figure}
{
\centering
\includegraphics[width=0.7\linewidth]{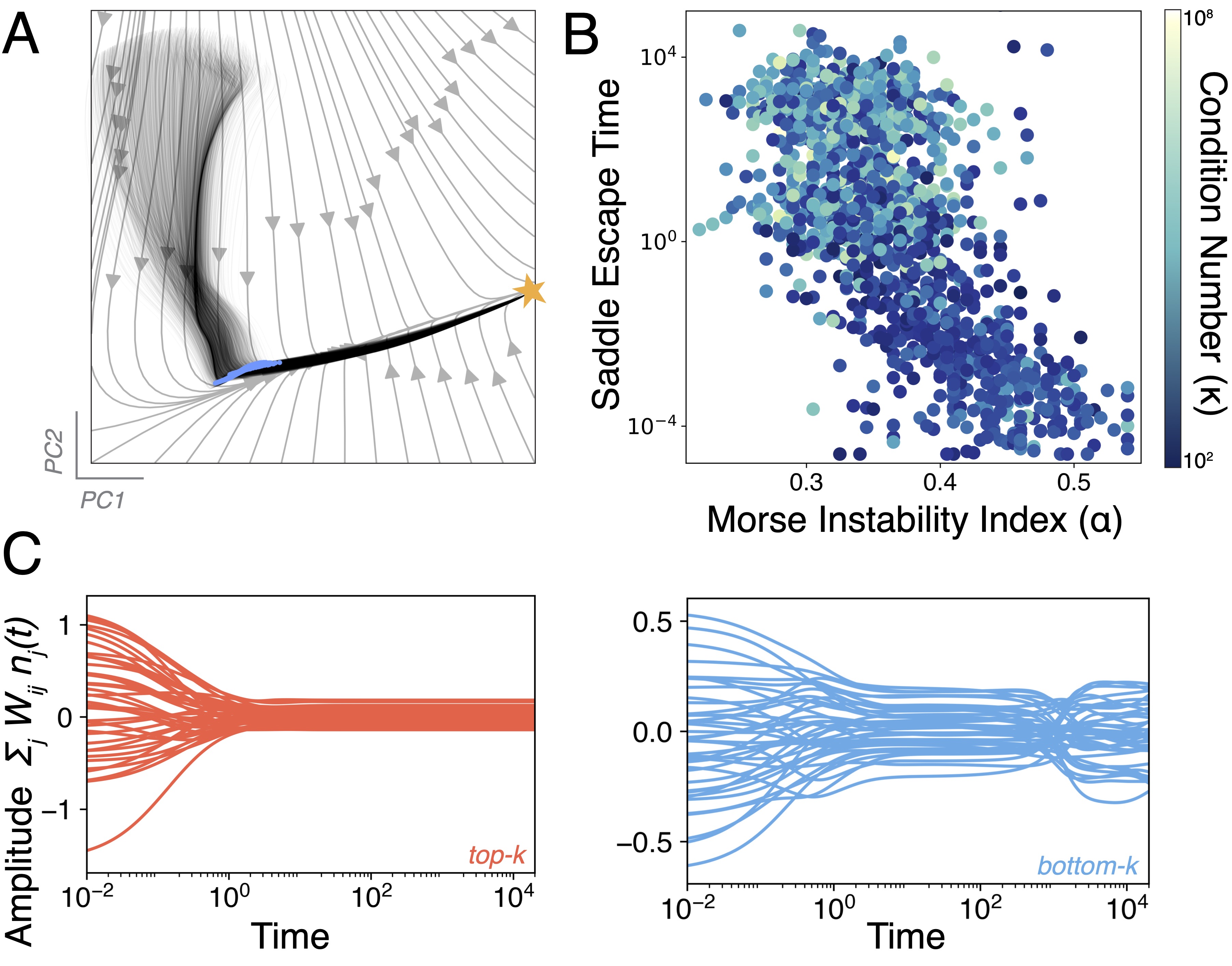}
\caption{
\textbf{Slow manifolds form a complex optimization landscape.} 
(A) An embedding of $10^3$ trajectories with different initial conditions in an ill-conditioned ecosystem ($\epsilon > 0$). The global equilibrium is marked with a star, and the corresponding solutions of the degenerate ($\epsilon = 0$) case are overlaid (blue). 
(B) Time that ill-conditioned trajectories for different random ecosystems (colored by condition number) spend near the former ($\epsilon=0$) solutions, versus the solution's Morse instability index.
(C) Projection of a single trajectory onto the right singular vectors associated with the largest (red) and smallest (blue) singular values. 
}
\label{precon}
}
\end{figure}

Given a high-dimensional experimental time series of species abundances, dimensionality reduction techniques are frequently used to map the dynamics to reduced-order coordinates \cite{tikhonov2017theoretical}. For example, when principal components analysis is applied to time series from a microbial ecosystem, the leading principal components correspond to groups of dynamically-correlated species that serve equivalent functions, such as producers, consumers, or scavengers \cite{tikhonov2017theoretical,frentz2015strongly}. These groups of strongly-correlated species, termed \textit{ecomodes}, identify species that act in concert, and thus reveal functional redundancy directly from observed dynamics.

We apply these methods to a sample of trajectories from an $N=10^3$-dimensional ill-conditioned ecosystem by embedding the dynamics into two dimensions using principal components analysis (Fig. \ref{precon}A, black traces). We find that trajectories originating from diverse initial conditions become trapped along a low-dimensional manifold. Setting $\epsilon = 0$ and then reintegrating each trajectory reveals that this manifold comprises a set of degenerate equilibria, which becomes a trapping region when $\epsilon > 0$ (Fig. \ref{precon}A, blue points). Gradual traversal of this slow manifold generically produces long transients; trajectories approach the global fixed point only after escaping the slow manifold. Slow transients due to weakly-unstable solutions have previously been characterized in complex foodwebs \cite{hastings2018transient}, and they particularly arise in systems with cyclic dominance and succession \cite{morozov2020long}. These slow manifolds represent a subspace within the full $N$-dimensional phase space on which intragroup dynamics unfold. For example, if one of two nearly-redundant nitrogen-fixing species in a microbial mat has a slightly higher survival rate, its gradual exclusion of the less-fit species occurs along the slow manifold. In numerical analysis, optimization of random high-dimensional landscapes is often dominated by saddle points, which become exponentially more likely than local minima as problem dimensionality grows \cite{dauphin2014identifying}. Recent theoretical results confirm this effect for \eqref{lv} in the well-conditioned ($P=I$) limit: as the number of species grows, the system spends more time trapped near unstable equilibria \cite{ben2021counting,altieri2021properties,bunin2017ecological,biroli2018marginally}. We thus attribute the empirical success of dimensionality reduction for ecological datasets to the existence of slow manifolds, which represent low-dimensional trapping regions within the full $N$-dimensional space in which the dynamics unfold.

\begin{figure*}
{
\centering
\includegraphics[width=0.9\linewidth]{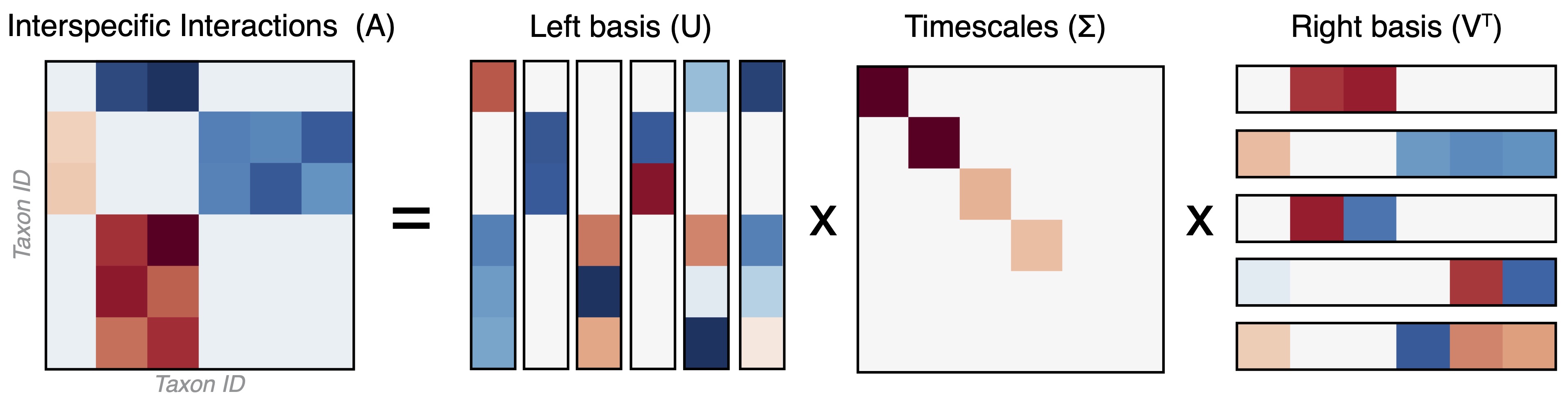}
\caption{
\textbf{Singular value decomposition of a species interaction matrix.} 
A schematic of singular value decomposition of the hierarchical species interaction matrix shown in Figure \ref{fig:model}. The left and right sets of singular vectors $U, V$ isolate groups of species that are functionally redundant, while the diagonal elements $\sigma_i$ in the singular value matrix $\Sigma$ encode the hierarchy of timescales that emerge due to functional redundancy.
}
\label{fig:svd}
}
\end{figure*}

Because our formulation \eqref{lv} contains slow manifolds induced by the symmetry-breaking perturbation $\epsilon$, we consider the degree that these manifolds globally shape the dynamics. Recent work on low-rank structure in large datasets ties low effective dimensionality to the structure of underlying interaction networks \cite{thibeault2024low}. Any interaction matrix $A$ admits a real-valued singular value decomposition of the form $A = U \Sigma V^\top$, where the diagonal matrix $\Sigma \in \mathbb{R}^{N \times N}$ contains the singular values $\sigma_i$ of $A$ in decreasing order along its main diagonal (Fig. \ref{fig:svd}) \cite{trefethen2022numerical}. Physically, this decomposition isolates groups of related species within the basis sets $U,V \in \mathbb{R}^{N \times N}$ that fluctuate characteristic timescales $\sigma_i$.
For our interaction matrix \eqref{interact}, the magnitudes of the singular values, and thus relevant timescales in the system, obey Weyl's inequality $|\sigma_i(A) - \sigma_i(A - \epsilon E) | \leq \epsilon \sqrt{\rho \, N}$. This implies that $A$ is a low-rank matrix with $N - M$ singular values $\sim \epsilon$ \cite{thibeault2024low}. As a result, the $N$-dimensional dynamics rapidly converge onto $M$-dimensional slow manifolds. These slow modes (formerly stable solutions in the $\epsilon = 0$ case) are misaligned across phase space, leading to a separation between gradual dynamics on slow modes, and fast high-dimensional transitions among them. On a given mode $\v{n}^*$, we perform a matched expansion of \eqref{lv} in terms of $\epsilon$ and a small perturbation off the slow manifold $\v{n}(t) = \v{n}^* + \delta \v{n}(t)$, producing the linearized dynamics $\dot{\delta n_i}(t) = n_i^* \sum_j (A_{ij} \delta n_j(t) + \epsilon E_{ij} {n}_j^*)$. The first term represents linear dynamics governed by the Jacobian of the unperturbed system ($J_{ij} \equiv n_i^* A_{ij}$), commonly known as the community matrix \cite{hastings2010regime}. The second term prevents equilibration by inducing slow $\sim\!\epsilon$ drift along the manifold. To quantify this kinetic trapping effect, we numerically identify slow manifolds for each of the random ecosystems shown in Fig. \ref{transient}B, and measure its relative stability via the scaled Morse instability index, $\alpha \equiv N^- / N$. The Morse index quantifies the relative number of unstable "escape" directions $N^- \leq N$ emanating from a given saddle region $\v{n}^*$ based on the eigenvalue spectrum of $J$. Saddles with relatively few downwards directions ($0 < \alpha \ll 1$) generally take longer to escape, leading to a smaller Morse index $\alpha$. We confirm this effect by directly measuring the escape time for each pseudosolution (Fig. \ref{precon}B). The escape time linearly decreases with $\alpha$ across different ecosystems, consistent with the theoretical expectation for hyperbolic dynamical systems \cite{kantz1985repellers}. 

To better understand dimensionality reduction and ecomode decomposition of ecological time series, we separately study the dynamics associated with fast and slow modes. 
Rather than directly study the full $N$-dimensional dynamics $\v{n}(t)$, we perform a linear transformation $W \in \mathbb{R}^{N \times K}$ that projects the dynamics into a $K \leq N$ dimensions. 
Recent results for low-rank systems show that the particular projection $W = \hat{V}^\top$ minimizes the least-squares alignment error between the vector field of a high-dimensional system and its $K$-dimensional reduced-order projection. The columns of $\hat{V}$ contain the first $K$ right singular vectors $V$ in the singular value decomposition of $A$ \cite{thibeault2024low}.
Because the matrix $A$ has $N$ right singular vectors, we choose two projections $W_\text{fast} = V_{:,1:K} \in \mathbb{R}^{N \times K}$ and $W_\text{slow} = V_{:,(N-K+1):N} \in \mathbb{R}^{N \times K}$, and apply them to an ill-conditioned ecosystem's trajectory in Fig. \ref{precon}C.

We find that projection onto the ecomodes ($W_\text{fast} \v{n}(t)$ and $W_\text{slow} \v{n}(t)$) separates fast timescales for relaxation onto the slow manifold network, and slow ($\sim \kappa$) timescales associated with navigating these modes (Fig. \ref{precon}C). In numerical optimization theory, ill-conditioning is mitigated by finding a \textit{preconditioning transformation} $-W A \v{n} = W \v{r}$ such that $\kappa(W A) \ll \kappa(A)$. For our system, we observe that both $W_\text{fast}$ and $W_\text{slow}$ effectively precondition the dynamics, with $\kappa(W_\text{fast} A)/\kappa(A) < 10^{-4}$ and  $\kappa(W_\text{slow} A)/\kappa(A) < 10^{-5}$. In general, the condition number $\kappa(A)$ may always be expressed as a ratio of the extremal singular values of $A$,
\begin{equation}
    \kappa(A) = \sigma_\text{max}(A) / \sigma_\text{min}(A)
\label{cond_sing}
\end{equation}
implying that the condition number essentially measures timescale separation in \eqref{lv}. Applying dimensionality reduction to the dynamics of a complex ecosystem therefore preconditions by isolating singular values associated with fast and slow dynamics. Preconditioning is known to isolate fast and slow subspaces in iterative linear solvers \cite{trefethen2022numerical}, and our findings connect this phenomenon to the empirical success of ecomode analysis in describing high-dimensional foodwebs, especially microbial ecosystems \cite{frentz2015strongly,tikhonov2017theoretical,louca2016high}. When $A$ is symmetric, our preconditioning argument supports prior works that construct ecomodes using eigenvectors of the interaction matrix \cite{tikhonov2017theoretical}. These works also find slow manifolds associated with degenerate eigenvalues, which identify substitutable species and thus the rank (effective size) of the ecosystem \cite{tikhonov2017theoretical}. However, because the eigenvectors of a non-symmetric matrix have no general relationship with the right singular vectors, the projection $V^\top$ should be preferred for systems with non-reciprocal (e.g. predator-prey or consumer-resource) interactions.

\subsection{Transient chaos slows approach to equilibrium.}

\begin{figure*}
{
\centering
\includegraphics[width=0.7\linewidth]{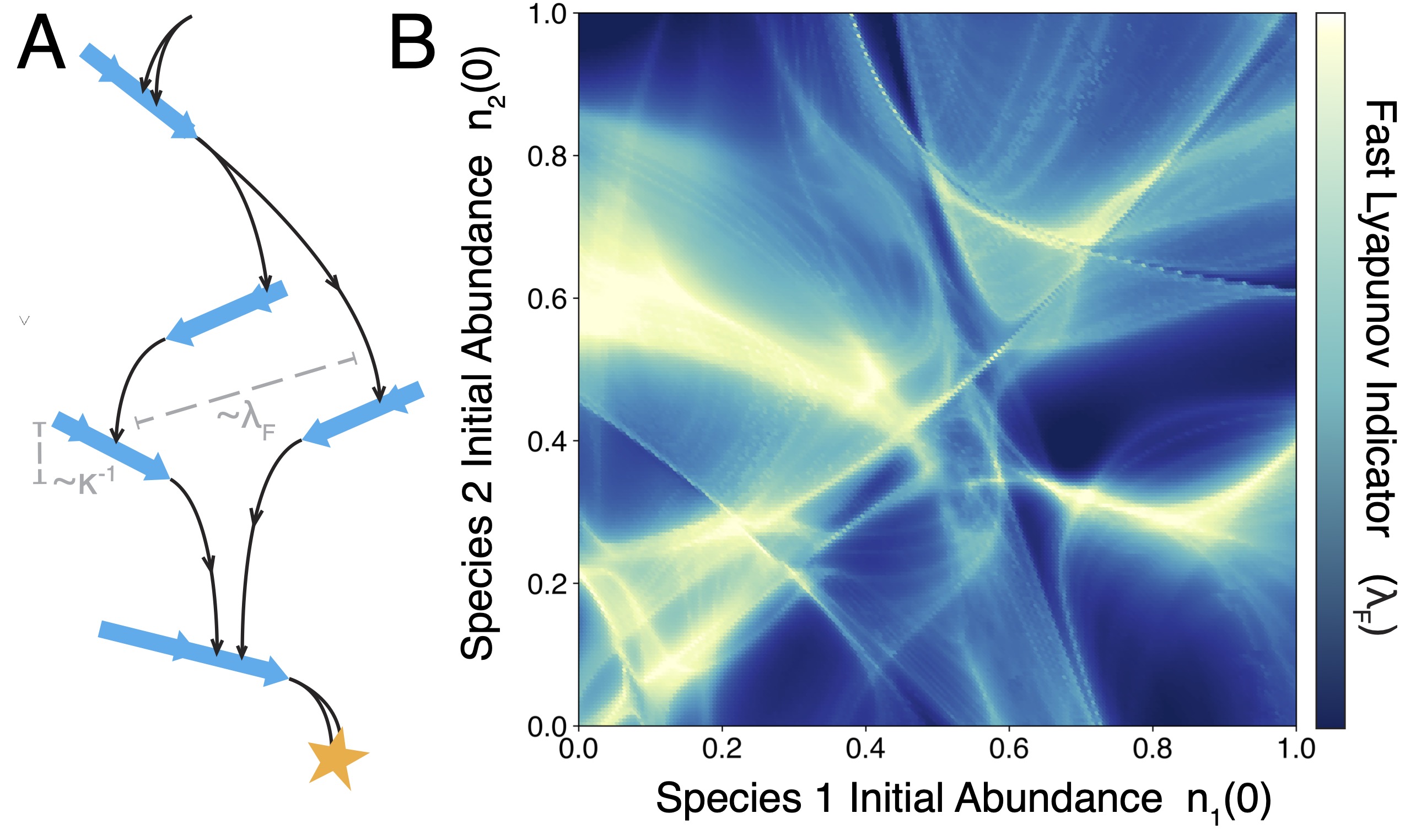}
\caption{
\textbf{Transient chaos due to slow manifold scattering.} (A) The "pachinko" mechanism for ill-conditioned dynamics, in which slow manifolds temporarily disperse neighboring trajectories that later reunite at the global equilibrium. (B) Caustics in the Fast Lyapunov Indicator ($\lambda_F$) versus initial conditions on a two-dimensional slice through the $N$-dimensional space of initial species densities. 
}
\label{basins}
}
\end{figure*}

Random ill-conditioned ecosystems are harder to solve due to the need to navigate a network of slow manifolds, suggesting that such systems pose more computationally-demanding optimization problems. But where is this excess computation allocated? The physical Church-Turing thesis asserts that continuous simulation of a hard combinatorial optimization problem likely incurs exponential cost \cite{vergis1986complexity}. Recent continuous formulations of discrete constraint satisfiability problems show that hard problem instances exhibit transient chaos, which manifests as exponentially-increasing sensitivity (and thus required precision) and fractal basin boundaries across initial conditions \cite{ercsey2011optimization}. For our system, we surmise that the slow manifolds act analogously to constraints, producing an irregular landscape of slow barriers that temporarily scatter trajectories as they approach the global fixed point, a mechanism reminiscent of a pachinko game (Fig. \ref{basins}A). To investigate this effect, we compute the fast Lyapunov indicator ($\lambda_F$), a quantity originally developed by astrophysicists to quantify the stability of planetary orbits in high dimensions \cite{froeschle2000graphical}. For each trajectory $\mathbf{x}(t)$ of \eqref{lv}, we integrate the variational equation $\dot{\mathbf{w}}(t) = \mathbf{J}[\mathbf{x}(t)] \mathbf{w}(t)$, where $\mathbf{w}(0) = I \in \mathbb{R}^{N \times N}$. The quantity $\lambda_{F} = \max_{t} \log\| \mathbf{w}(t) \|_2$ represents the maximum chaoticity encountered on a trajectory, a quantity more informative than traditional Lyapunov exponents for transient dynamics where $\lim_{t \rightarrow \infty}\log\| \mathbf{w}(t) \|_2 / t \leq 0$. We note that the dynamics $\dot{\mathbf{w}}(t)$ represent a continuous-time variant of the power method, a standard iterative technique for finding the largest eigenvalues of square matrices \cite{trefethen2022numerical}. We therefore interpret $\lambda_{F}$ as a spatially-resolved probe of the local conditioning of $A$ for a given initialization $\v{n}(0)$. 

A random two-dimensional slice through the space of initial species densities reveals intricate pseudobasins, indicating abrupt changes in the route by which the ecosystem approaches equilibrium when initial species densities change by small amounts (Fig. \ref{basins}B). These patterns resemble optical caustics, which arise when density fluctuations in a transparent medium distort the pathlengths of light rays. Here, the slow manifolds play a similar role by intermittently trapping and scattering trajectories arising from different initial conditions. This slow-mode mechanism differs from the scrambling induced by chaotic saddles in high-dimensional open chaotic systems like planetary orbits \cite{froeschle2000graphical,lai2011transient}, in which conservation of mechanical energy precludes approach to a fixed point. Rather, we find that \eqref{lv} exhibits doubly-transient chaos, a recently-characterized phenomenon in dissipative systems that approach global fixed points by differing routes \cite{motter2013doubly}. Although a globally-stable community will ultimately approach the same equilibrium regardless of its assembly route or initial species densities, transient chaos introduces contingency because minute fluctuations cause the time to reach equilibrium to vary by three orders of magnitude.

\subsection{Selection for diversity creates ill-conditioned ecosystems.}

\begin{figure*}
{
\centering
\includegraphics[width=0.7\linewidth]{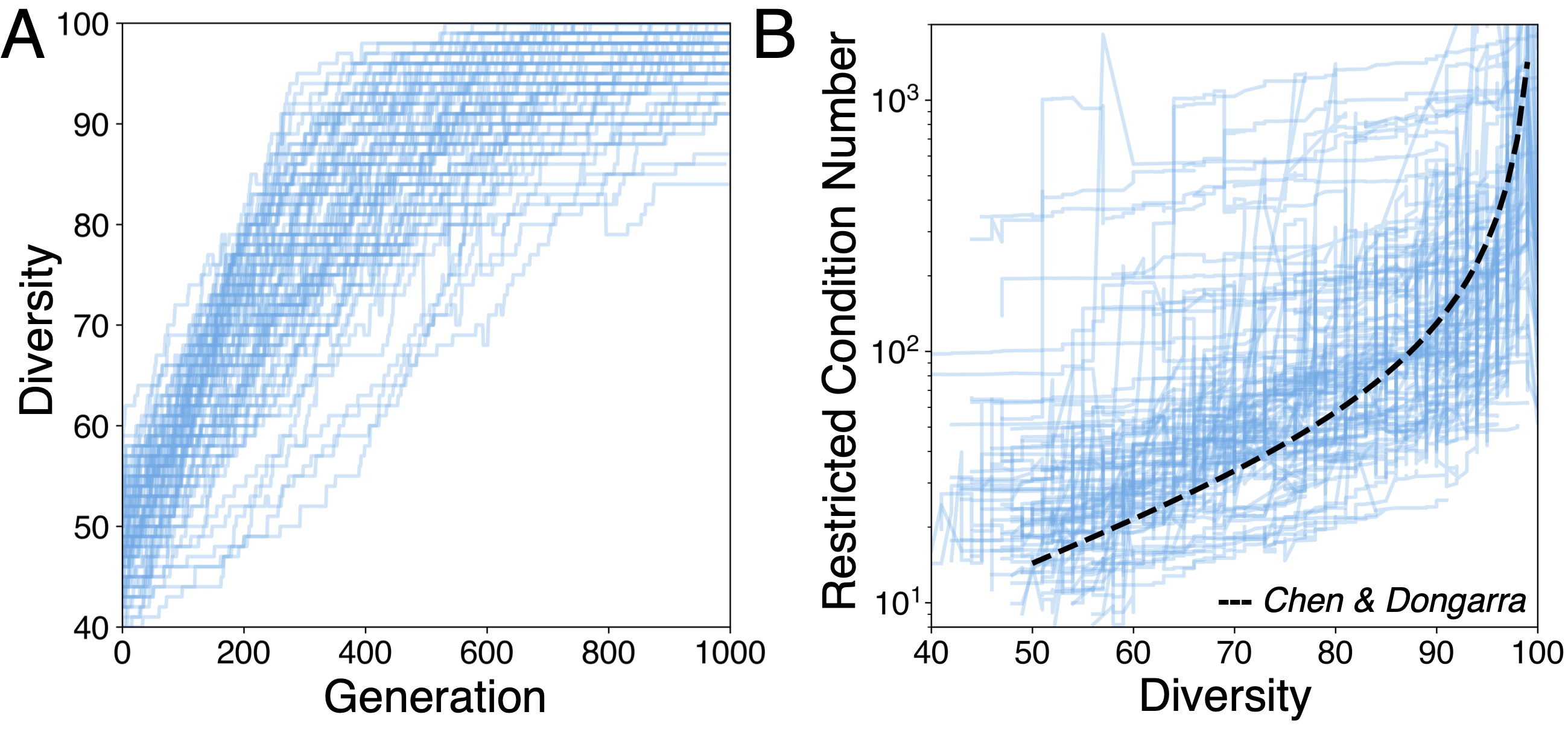}
\caption{
\textbf{Selection for diversity produces ill-conditioning.} (A) The condition number $\kappa$ versus generation for $10^3$ replicate random ecosystems of $N=10^3$ species evolved to have high steady-state diversity, defined as the number of coexisting species at equilibrium. (B) The restricted condition number $\hat\kappa$ of the interaction matrix of species that survive at steady-state, versus the overall steady-state diversity. Dashed line indicates expected condition number for a random $N \times N_\text{coex}$ matrix with normally-distributed elements.
}
\label{evolve}
}
\end{figure*}

We next consider whether ill-conditioned dynamics can occur in real ecosystems, which typically have more structured interaction networks than \eqref{interact}. Using a genetic algorithm, we evolve random foodwebs by selecting for an increased number of surviving species at steady-state, $N_\text{coex}$, a measure of diversity in May's original work \cite{may1972will}. In our procedure, we start by randomly sampling $10^{3}$ distinct well-conditioned random foodwebs ($P = I$ in \eqref{interact}), and then simulate their dynamics until they reach equilibrium $\v{n}^*$. We then select the 10\% of foodwebs with the highest diversity $N_\text{coex} \equiv \| \v{n}^* \|_0$. We propagate these foodwebs to the next generation, and cull the remaining $90\%$ and replace them with new foodwebs assembled from random blocks of row-column pairs sampled from the surviving $10\%$. We repeat this procedure for $10^3$ generations.

We find that evolving foodwebs towards higher diversity $N_\text{coex}$ robustly produces higher-$\kappa$ ecosystems, independently of modelling choices like foodweb scale or mutation and crossover rates (Fig. \ref{evolve}A). The robust emergence of ill-conditioning implies that an intrinsic physical tradeoff, rather than a dynamical effect, produces the observed scaling. Intuitively, as the diversity $N_\text{coex}$ increases, fewer positivity constraints in \eqref{lv} remain active at equilibrium, and so the system becomes more linear. 
Additional coexisting species increases the probability that a subset serve a substitutable role, leading to a higher condition number. This effect occurs in experimental microbial ecosystems, in which substitutable consortia of closely-related species emerge at high diversities \cite{goldford2018emergent,louca2016decoupling}. To measure this effect, we introduce a \textit{restricted condition number} $\hat{\kappa}$ describing the minor matrix comprising only the columns of $A$ associated with nonzero entries in $\v{n}^*$. Principal minors of a positive definite matrix remain positive definite, and so the reduced system retains a global fixed point, corresponding to a noninvasible equilibrium of the reduced ecosystem containing only the species that survive at equilibrium. Computing $\hat\kappa$ for random ecosystems with varying $N_\text{coex}$ confirms a general relationship (Fig. \ref{evolve}B), which closely follows a known scaling law for the condition number of a random rectangular matrix, $\hat\kappa = 9.563 \,N/(N - N_\text{coex} + 1)$ (dashed line) \cite{chen2005condition}. This suggests that ill-conditioning arises primarily from selection for diversity, and not fine-tuned interaction values. Physically, this effect represents species packing: adding new species to an ecosystem increases the probability of duplicating an existing function. From an optimization standpoint, there are two competing effects: as $N_\text{coex}$ increases, fewer non-negativity constraints remain active, but the probability that two species are redundant increases. This interplay is known as complementary slackness in constrained optimization \cite{trefethen2022numerical}. Consistent with this interpretation, recent works map consumer-resource models (which generalize \eqref{lv}) onto quadratic optimization programs \cite{mehta2019constrained} and constraint satisfaction problems \cite{altieri2019constraint}, and find that niche overlap may be understood as redundancy emerging at high constraint loads. To confirm this connection, in the Appendix we repeat our analysis for a previously-studied variant of \eqref{lv} that exhibits a controllable transition from multistability to global noninvasible stability \cite{rieger1989solvable,bunin2017ecological,ros2023generalized}. We observe a rapid increase in the conditioning $\kappa(A)$ near the critical point, as well as transient chaos. 
Taken together, these results suggest that, under generic selection for high steady-state diversity, ecosystems will exhibit ill-conditioning because functional redundancy becomes inevitable. Recent works describe species packing in crowded environments like the microbiome \cite{macarthur1970species,goldford2018emergent}, and our work shows that such systems likely exhibit long transients and wide variation in the timescales over which different subcommunities equilibrate.

\section{Perturbations experimentally probe the slow manifold}

\begin{figure*}
{
\centering
\includegraphics[width=0.7\linewidth]{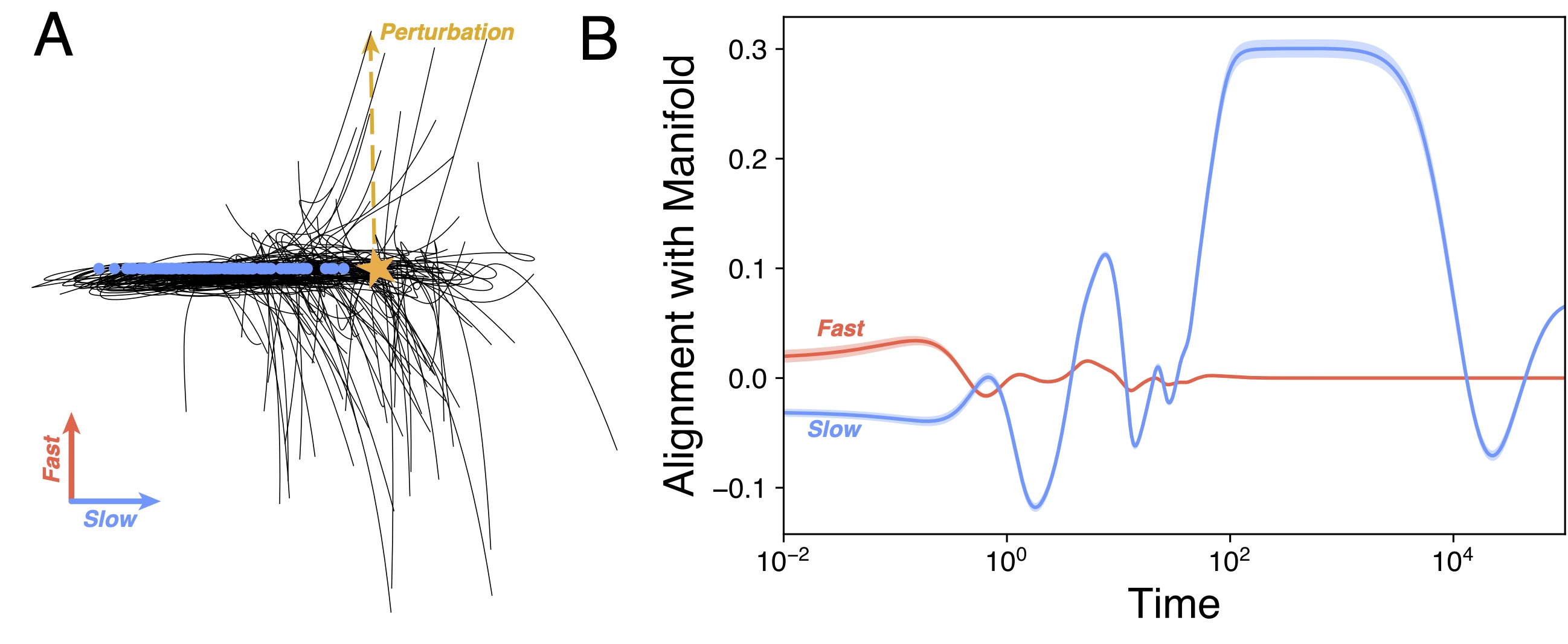}
\caption{
\textbf{Community response to perturbations reveals the slow manifold.} 
(A) A set of $100$ experiments in which a random pulse perturbation is applied to an equilibrated ecosystem, and the dynamics are allowed to return to steady state. The full $N = 200$ dimensional dynamics are projected onto the fastest and slowest directions associated with the singular vectors of the interaction matrix $A$. The slow manifold (blue points) is numerically detected by repeating the experiment with $\epsilon=0$ in \eqref{lv}.
(B) The Pearson correlation (inner product) between the velocity vector of the dynamics at each timepoint, and the fast and slow manifolds. Error bars represent standard errors over $100$ replicate communities.
}
\label{fig:perturb}
}
\end{figure*}

We next consider the detection of ill-conditioned dynamics in experimental settings. Testing our results for scaling laws and pseudobasins requires the ability to perform many replicate experiments with slight variations in initial conditions or parameter values, which is impractical at the current scale and precision of microbial ecosystem experiments \cite{louca2018function}. Moreover, it is difficult to fit an analytic model like \eqref{lv} directly to experimental data, which requires resolving $\sim\mathcal{O}(N^2)$ interaction parameters in a model of the form \eqref{lv}. 

However, we note that our framework is based on singular perturbation theory in numerical analysis. We therefore look to these methods for potential routes for measuring the conditioning $\kappa$ in real-world ecosystems, without the need to directly measure $A$ in \eqref{lv}. In numerical analysis, Krylov algorithms probe the properties of large matrices without directly materializing them \cite{trefethen2022numerical}. Two common approaches, power and inverse iteration, respectively estimate the largest and smallest eigenvalues of a large matrix $A$ by repeatedly applying a linear transformation derived from $A$ to a random vector. Because the condition number may be alternatively formulated as a ratio of singular values (\eqref{cond_sing}), we anticipate that, within a large ecological network, the properties of $A$ may be probed by applying a multifactorial "pulse" perturbation, in which multiple species are temporarily and simultaneously perturbed \cite{novak2016characterizing,stone2020stability}. In the non-invasible case, the ecosystem relaxes back to the global equilibrium $\mathbf{n}^*$, but the directions of fastest and slowest return to equilibrium will reveal the effective conditioning $\kappa$.

In Figure \ref{fig:perturb}, we apply $100$ random perturbations to an ecosystem of the form \eqref{lv} that is initialized at its global steady state $\v{n}^*$. Each perturbation corresponds to a random Gaussian vector of small amplitude $\delta \v{n} \sim \mathcal{N}(0, 10^{-3})$, though we set to zero any elements of the perturbation vector corresponding to species that are excluded at steady-state. The resulting dynamics always equilibrate back to steady state, but take widely-varying routes depending on the particular alignment of the random perturbation with slow manifolds of the underlying system.

We visualize the $N=200$-dimensional dynamics by projecting the trajectories onto the fastest and slowest modes, $\v{n}(t)\cdot \v{w}^{(N)} / \| \v{w}^{(N)} \|$ and $\v{n}(t)\cdot \v{w}^{(1)} / \| \v{w}^{(1)} \|$. These two directions correspond to the first and last right singular vectors of the interaction matrix $A$. While the interaction matrix $A$ is not directly measurable in real-world ecosystems, the extremal singular vectors correspond to principal components (or ecomodes) derived from trajectory time series \cite{frentz2015strongly}. In this projected basis, we observe that the dynamics quickly traverse the fast manifold by widely-varying routes, and then slowly approach the global equilibrium $\v{n}^*$ along the slow manifold.

However, our ensemble-based analysis is impractical in a strongly data-limited setting, such as a field experiment, in which only a single "pulse" perturbation occurs (such as a weather event or habitat change). We therefore construct a proxy for the slow and fast manifolds from a single time series (Fig. \ref{fig:perturb}B). Given a single time series, we estimate the velocity vector using finite differences. At long timescales after the pulse perturbation ends, this velocity vector becomes aligned (exhibits high Pearson correlation) with the direction of the slow manifold, consistent with emergent low-dimensionality in the observed dynamics. This direction remains consistent across different random perturbations, underscoring that the geometry of the slow manifold determines equilibration.
We thus find that slow, ill-conditioned dynamics associated with functional redundancy will produce a characteristic signature in an ecosystem's response to perturbations, even when an underlying model is unavailable.

\section{Discussion.}

Drawing on recent work in transient chaos and optimization theory, we have shown that equilibration of random ecosystems is globally constrained by ill-conditioning. 
Our work generalizes the classic stability vs. complexity dichotomy in mathematical biology \cite{may1972will}, by demonstrating how transients complicate approach to equilibrium. 
Physically, ill-conditioning corresponds to a case where in-group dynamics (e.g. competition among similar nitrogen-fixing species in a microbial mat) unfold over a slower timescale than intergroup competition, like community assembly from different niches \cite{gravel2016stability,pastore2021evolution}. 
This manifests as ultra-long transient dynamics, with durations that can vary by orders of magnitude given minor changes in the assembly sequence or initial species densities. 
Extended transients provide opportunities for noise, seasonal changes, climate variation, and other exogenous factors to disrupt or re-route the dynamics \cite{hastings2010regime}. 
Moreover, transient chaos implies that even an ecosystem that has successfully reached equilibrium is vulnerable to disruption, because transiently chaotic systems can undergo extended excursions away from equilibrium, if subjected to perturbations exceeding a finite threshold \cite{avila2023transition,novak2016characterizing}. 
Our results thus highlight that classical equilibrium-based analysis methods fail to fully characterize high-dimensional ecosystems, for which steady-state is an exception, not the rule.

We expect that our methods have relevance to other high-dimensional biological networks: the perfectly-substitutable case ($\epsilon \to 0$) implies that $A$ has $N-M$ repeated eigenvalues, a scenario equivalent to zero modes that arise in elastic models of allostery in biomolecules \cite{rocks2019limits}, as well as epistasis in genetic models \cite{husain2020physical}. Additionally, in statistical learning theory, high-dimensional optimization landscapes have been found to exhibit linear mode connectivity, in which networks of zero-barrier connections bridge between optima \cite{ainsworth2022git}, a concept extending earlier works describing symmetry-induced submanifolds that influence the dynamics of learning \cite{saad1995line}. These diverse systems all represent cases in which small perturbations may disrupt an intricate network of solutions, leading to slow, multiscale dynamics.

While our analysis largely relies on the particular form of \eqref{lv} as an analogue constrained linear regression, we emphasize that $\kappa$ physically represents the functional redundancy, and thus the computational cost to solve a linear system. 
Generalizations of \eqref{lv} instead sample $A$ from more structured probability distributions, in order to account for phenomena like niche competition or predation hierarchies \cite{tang2014reactivity,johnson2014trophic,gravel2016stability}. 
Other approaches use different dynamical equations, including stochastic models and empirical models learned directly from experimental data \cite{mehta2019constrained,poderoso2006controlling,wu2024data}.
Because all such models approach equilibrium, we expect that many of our observations are robust to the particular form of \eqref{lv}. This is because, in optimization theory, a generalized condition number exists for any constrained problem in terms of the distance to singularity within the feasible solution set \cite{renegar1995incorporating}. For example, in random Boolean satisfiability problems, the effective conditioning is governed by the ratio of constraints to variables \cite{ercsey2011optimization}. While direct calculation of the generalized condition number is prohibitive for many problem classes, its existence suggests that global computational constraints on equilibration exist for more complicated foodweb models, or other types of biological networks. Our work thus provides a tractable example of how intrinsic computational complexity may influence the organization of biological systems.

\begin{widetext}

\section{Table of Symbols}
\label{app:symbols}

\begin{table}[H]
\singlespacing
\centering
\begin{tabular}{>{\centering\arraybackslash}p{1in} p{5in}}
\hline
Symbol & Description \\
\hline
$N$ & The number of species. \\
$M$ & The number of functional trophic groups, $M \leq N$\\
$\epsilon$ & The scale of intraspecific variation within functional groups, equivalent to the amplitude of the singular perturbation. \\
$A$ & The $N \times N$ matrix of interactions among all species. \\
$P$ & The assignment matrix describing functional redundancies among groups, such as trophic levels. \\
$A^{(0)}$ & The random initial matrix of interactions among species, before functional redundancies are included.  \\
$r_i$ & The growth rate in isolation of the $i^{th}$ species \\
$E$ & The random matrix encoding minor variations among functionally redundant groups.\\
$\kappa$ & The condition number of the interaction matrix, which determines the ratio of the largest to the smallest relevant timescales in a community. \\
$\hat\kappa$ & The restricted condition number, corresponding to the condition number of the reduced matrix of interactions among only the species that survive at equilibrium. \\
${n}_i(t)$ & The instantaneous density of the $i^{th}$ species at a given time $t$ \\
$n_i^*$ & The densities of the $i^{th}$ species after the ecosystem reaches steady-state \\
$\tau$ & The equilibration time that it takes the ecosystem to reach steady-state \\
$N_\text{coex}$ & The number of surviving (coexisting) species at steady-state, $N_\text{coex} \leq N$ \\
$d$ & The intraspecific density limitation \\
$\rho$ & The connectance, or the number of all interactions in the ecosystem divided by the total number of possible interactions \\
$\alpha$ & The Morse instability index, indicating the relative rate of escape from a saddle point\\
$\delta n_i(t)$ & The instantaneous difference between the density of the $i^{th}$ species, and its value at steady-state\\
$\sigma_i(A)$ & The $i^{th}$ singular value of the matrix $A$, ranked in descending order of magnitude \\
$\lambda_F$ & The fast Lyapunov indicator, a measure of the relative sensitivity of the route to equilibrium associated with a given set of species densities\\
\hline
\end{tabular}
\caption{Mathematical symbols and their descriptions.}
\end{table}

\section{Code Availability}

Code for this study is available at \url{https://github.com/williamgilpin/illotka}

\section{Acknowledgements}
W.G. thanks Aleksandra Walczak, Rob Phillips, Yuanzhao Zhang, and Annette Ostling for informative discussions. 
WG was supported by NSF DMS 2436233 and NSF CMMI 2440490.
This project has been made possible in part by Grant No. DAF2023-329596 from the Chan Zuckerberg Initiative DAF, an advised fund of Silicon Valley Community Foundation.
Computational resources for this study were provided by the Texas Advanced Computing Center (TACC) at The University of Texas at Austin. 

\section{Supporting Information}

\noindent\textbf{Fig. \ref{altmodel}. Complex transients in an alternative ecosystem model.} (A) The condition number as an ecosystem model transitions from noninvasible global stability to multistability. Settling time $\tau$ versus condition number $\kappa$ for $10^{4}$ random communities. (B) Caustics in the Fast Lyapunov Indicator ($\lambda_F$) versus initial conditions on a two-dimensional slice through the $N$-dimensional space of initial species densities. 

\noindent\textbf{Fig. \ref{fig:connectivity2}. The effect of network connectivity on scaling of transient dynamics.} Settling time $\tau$ versus condition number $\kappa$ for $3 \times 10^{3}$ random communities; dashed line shows the scaling expected for an iterative linear program solver. Colors indicate communities with different values of the connectance ($\rho$).

\noindent\textbf{Fig \ref{fig:connectivity}. The effect of network connectivity on transient dynamics.} Caustics in the Fast Lyapunov Indicator ($\lambda_F$) versus initial conditions on a two-dimensional slice through the $N$-dimensional space of initial species densities. Panels correspond to three levels of network connectance ($\rho$).

\noindent\textbf{Fig. \ref{fig:rank_scaling}. The effect of slow manifold dimension on scaling of transient dynamics.} Settling time $\tau$ versus condition number $\kappa$ for $3 \times 10^{3}$ random communities; dashed line shows the scaling expected for an iterative linear program solver. Colors indicate communities with different slow manifold dimensionalities ($M/N$).

\noindent\textbf{Fig. \ref{fig:rank_basins}. The effect of  slow manifold dimension on transient dynamics.} Caustics in the Fast Lyapunov Indicator ($\lambda_F$) versus initial conditions on a two-dimensional slice through the $N$-dimensional space of initial species densities. Panels correspond to three values of the slow manifold dimensiona ($M/N$).


\clearpage
\renewcommand{\thetable}{S\arabic{table}}
\setcounter{table}{0}
\renewcommand{\thefigure}{S\arabic{figure}} 
\setcounter{figure}{0}
\renewcommand{\theequation}{A\arabic{equation}}
\setcounter{equation}{0}
\renewcommand{\thesubsection}{\Alph{subsection}}
\setcounter{subsection}{0}
\renewcommand{\thesection}{Appendix \Alph{section}}
\setcounter{section}{0}


\setcounter{page}{1} 
\section{Code Availability}

Code for this study is available at \url{https://github.com/williamgilpin/illotka}

\section{Slow modes in the Generalized Lotka-Volterra model.}

The generalized Lotka-Volterra model has the form
\begin{equation}
    \dot{n}_i(t) = n_i(t) \left( r_i + \sum_{j=1}^N A_{ij} n_j(t)  \right).
\label{lv_supp}
\end{equation}
The growth rates $r_i\sim \mathcal{N}(0, 1)$ include species capable of growing in isolation ($r_i > 0$, such as autotrophs) and those that require other species to survive $r_i < 0$. The matrix $A$ is sampled from the family of matrices
\begin{equation}
    A = P^T (A^{(0)} - d\, I) P + \epsilon\,E,
\label{interact_supp}
\end{equation}
where $A^{(0)}_{ij}, E_{ij} \sim \mathcal{N}(0, 1)$, $\epsilon \ll 1$, $d$ is a density-limitation constant, and $I$ is the identity matrix. We introduce the notation $A = A^{(0)} + \epsilon \,E$, where $A^{(0)} \equiv P^T (A^{(0)} - d\, I) P$. We linearize \eqref{lv_supp} around a slow manifold $\v{n}^*$, such that $\v{n}(t) = \v{n}^* + \delta \v{n}(t)$ and $- A^{(0)} \v{n}^* = \v{r}$. Matching terms of equivalent order results in linearized dynamics
\begin{equation}
    \dot{\delta n}_i(t) = n_i^* \sum_{j=1}^N A^{(0)}_{ij} \,\delta n_j(t) + \epsilon\, n_i^* \sum_{j=1}^N E_{ij} \,n_j^*
\label{pert}
\end{equation}
where we have discarded terms $\sim\!\mathcal{O}(\delta n_i \delta n_j)$ and $\sim\!\mathcal{O}(\epsilon \delta n_i)$. We note that retaining the latter terms would lead to perturbations $\sim\!\mathcal{O}(\epsilon)$ to the eigenvalues of the unperturbed Jacobian; however, in the globally-stable regime, the dynamics near the equilibrium point are dominated by the density-dependent term $d$, and so these cross-terms have minor effects on the global fixed point. 

The first term in \eqref{pert} creates linearized dynamics described by the unperturbed Jacobian matrix (community matrix) \cite{novak2016characterizing}. The unperturbed Jacobian matrix may be alternatively written as the matrix product $A\; \text{diag}(\v{n}^*)$, which is independent of $\epsilon$. The second term biases the dynamics away from the unperturbed fixed point $\v{n}^*$, creating a flow with speed $\sim\!\epsilon$ along the slow manifolds.

\section{Numerical Integration.}
\label{app_integrate}

All numerical results are generated with an implicit embedded Runge-Kutta solver (Radau) that inverts the analytic Jacobian at each timestep. Relative and absolute tolerances of the solver are set to $<10^{-12}$, ensuring accuracy comparable to recent studies of transient chaos \cite{zhang2023catch,motter2013doubly}. 

For calculation of the settling time, we numerically integrate until $\dot{\v{n}}(t) < 10^{-14}$, and we then use the eigenvalues of analytical Jacobian of \eqref{lv_supp} to confirm that the asymptotic value is a local minimum, which we consider an estimate of the global solution $\v{n}^*$. We then use a nonnegative least-squares solver (the Lawson-Hanson algorithm implemented in \texttt{scipy.optimize.nnls}) to directly verify that $-A \v{n}^* = \v{r}$. We then retroactively find the settling time $\tau$ by choosing the fixed convergence floor $\xi = 10^{-7}$ and find $\tau \equiv \inf_t (\| \mathbf{n}(t) - \mathbf{n}^* \|_2 \leq \xi)$. The precision floors are chosen so that, in all cases, $\tau$ occurs before the termination of integration.

\section{Calculation of the Fast Lyapunov Indicator.}

For calculation of the Fast Lyapunov Indicator (FLI), the variational equation $\dot{\v{w}}(t) = J[\v{n}(t)] \v{w}(t)$, $\v{w}(0) = I$ is introduced based on the analytic Jacobian of \eqref{lv_supp}. The variational equation is integrated concurrently with the original trajectory, with all timesteps controlled by the dynamics of $\v{n}(t)$. During integration, whenever $\| \v{w}(t) \|_2$ grows too large for numerical stability, we rescale all matrix elements and then store the scale factor, allowing the exact value of $\| w(t) \|_2$ to be reconstructed \textit{post hoc}. For a given initial condition $\v{n}(0)$, the Fast Lyapunov Indicator is given by
\[
    \lambda_{F} = \max_{t} \log\| \v{w}(t) \|_2
\]
where the $\v{w}(t) \in \mathbb{R}^{N \times N}$ and $\v{w}(0) = I$ \cite{froeschle1997fast}. The variational equation may be interpreted as evolving each column of $\v{w}(t)$ separately. Because each column points in a different initial direction, the columns stretch at different rates depending on whether they happen to initially align  with the dominant stretching directions of $J(\v{n}(t))$. At long times, the fastest-growing column dominates the matrix norm. $\lambda_{F}$ may thus be interpreted as the fastest exponential divergence observed at any time, and along any initial direction, for a given point $\v{n}(0)$ in the domain of a dynamical system. This property contrasts $\lambda_{F}$ with the true Lyapunov exponent $\lim_{t \rightarrow \infty}\log\| \v{w}(t')\| / t$, which eventually approaches zero or a negative value for most initial conditions in systems that exhibit scattering or a global fixed point. Prior works have shown that $\lambda_{F}$ can robustly detect transient events unfolding over a range of timescales, including short-period scrambling events in the restricted three-body problem \cite{lega2011detection}, long-timescale resonances in the solar system \cite{todorovic2020arches,fouchard2002relationship}, and gradual diffusion in quasi-integrable Hamiltonian systems \cite{froeschle2000graphical}.

\section{Embedding the slow manifold dynamics.} We fix the parameters of an ecosystem \eqref{lv_supp}, with $\v{F}(\v{n}; A, \v{r})$ denoting the right hand side of the differential equation. We compute a set of $N_\text{traj}$ simulations originating from different initial conditions and sampled at $T$ discrete timepoints, $\v{X}_\text{traj} \equiv \{\v{n}^{(k)}(t)\} \in \mathbb{R}^{N_\text{traj} \times T \times N}$. On this set of trajectories, we perform principal components analysis (PCA) and transform the data into the full principal components space $\v{Y}_\text{pca} \in \mathbb{R}^{N_\text{traj} \times T \times N}$. For visualization purposes, we truncate this matrix in order to view the trajectories along only the top two principal components, $\hat{\v{Y}}_\text{pca} \in \mathbb{R}^{N_\text{traj} \times T \times 2}$. For each initial condition, we perform a replicate integration with $\epsilon=0$ in order to identify the set of fixed points for the degenerate case, $\v{X}_\text{traj}^{(0)*} \in \mathbb{R}^{N_\text{traj} \times N}$. We project these unperturbed solutions into the truncated PCA coordinates, in order to visualize in low dimensions the locations of formerly-stable solutions, $\hat{\v{Y}}_\text{pca}^{(0)*} \in \mathbb{R}^{N_\text{traj} \times 2}$.

To obtain a vector field and streamlines, we define a uniform mesh over the first two PCA embedding coordinates $\hat{\v{Y}}_\text{mesh} \in \mathbb{R}^{N_\text{mesh} \times 2}$, which we then promote to a full-dimensional mesh by appending the mean of the remaining $(N-2)$ PCA coordinates, producing the coordinates $\v{Y}_\text{mesh} \in \mathbb{R}^{N_\text{mesh} \times N}$. We then invert the PCA transformation (a linear transformation and a shift by the featurewise mean vector) in order to pull $\v{Y}_\text{mesh}$ back into the ambient coordinates $\v{X}_\text{mesh} \in \mathbb{R}^{N_\text{mesh} \times N}$. In the inverted mesh coordinates, we evaluate \eqref{lv_supp} at each mesh location to produce a velocity vector field $\v{U} = \v{F}(\v{X}_\text{mesh}; A, \v{r}) \in \mathbb{R}^{N_\text{mesh} \times N \times N}$ in the ambient coordinates, which we then project back into the truncated PCA basis $\hat{\v{V}} \in \mathbb{R}^{N_\text{mesh} \times 2 \times 2}$ in order to visualize the vector field and streamlines along the slow manifold. Trajectories in the embedded coordinates appear to cross streamlines, because the observed dynamics are a projection of higher-dimensional trajectories from a velocity field $\v{V} \in \mathbb{R}^{N_\text{mesh} \times N \times N}$, where streamlines do not cross.

\section{Evolutionary simulations lead to ill-conditioned dynamics}
\label{app_genetic}

We randomly sample $10^3$ random interaction matrices with $P=I$, $\epsilon=0$ in \eqref{interact_supp}. We integrate their dynamics under \eqref{lv_supp}, using the same integration precision and termination conditions described in \ref{app_integrate}. Consistent with prior results \cite{servan2018coexistence}, we observe that the initial steady-state diversity, $N_\text{coex} \equiv \| \v{n}^* \|_0$, exhibits a binomial distribution across replicates, with $\langle N_\text{coex} \rangle_A \approx N / 2$.

We next select the top $10\%$ of foodwebs ($A$ matrices) to survive to the next generation, and we delete the remaining $90\%$. We replace these culled foodwebs with new foodwebs with interactions randomly sampled from the surviving foodwebs. However, we hold constant the vector $\mathbf{r}$ of growth rates for each species. This is because the individual growth rates are the only distinguishing characteristic that would allow for distinct niches among the different species; if growth rates also evolve during the simulation, the system will trivially collapse onto a degenerate space of identical species in equal proportions. This case would be consistent with our observation that functional redundancy (and thus ill-conditioning) emerges at high diversity, but it otherwise provides little mechanistic insight.

We have also considered variants of our evolutionary procedure, including the addition of random mutations, constraints on crossover due to clamped values of the sparsity and norm of each $A$, and a more gradual recombination procedure, in which low-fitness ecosystems are not culled, but rather only partially updated based on interactions found in high-diversity ecosystems. While these hyperparameters affect evolution rate and convergence time, they have no effect on the scalings and tradeoffs we observe, suggesting that the phenomenon represents an intrinsic property of the space of ecosystems and $A$ matrices, rather than an artifact of our particular matrix sampling procedure.

From a biological perspective, our approach represents a limiting case of eco-evolutionary dynamics, in which population dynamics proceed rapidly compared to evolutionary dynamics. Interestingly, while the population dynamics represent an optimization in the sense of solving a non-negative least-squares problem, the process of evolving the $A$ matrices with genetic algorithms represent a higher-level meta-optimization,
\[
\begin{aligned}
    & \underset{A}{\text{maximize}}
    & & \|\mathbf{n}^*\|_0 \\
    & \text{subject to}
    & & -A \mathbf{n}^* = \mathbf{r}, \\
    &&& n_i^* \geq 0, \; \forall i.
\end{aligned}
\]
where the population dynamics are now encoded within an equality constraint.

\section{Alternative models.}
\label{app_alternates}

\begin{figure}
{
\centering
\includegraphics[width=\linewidth]{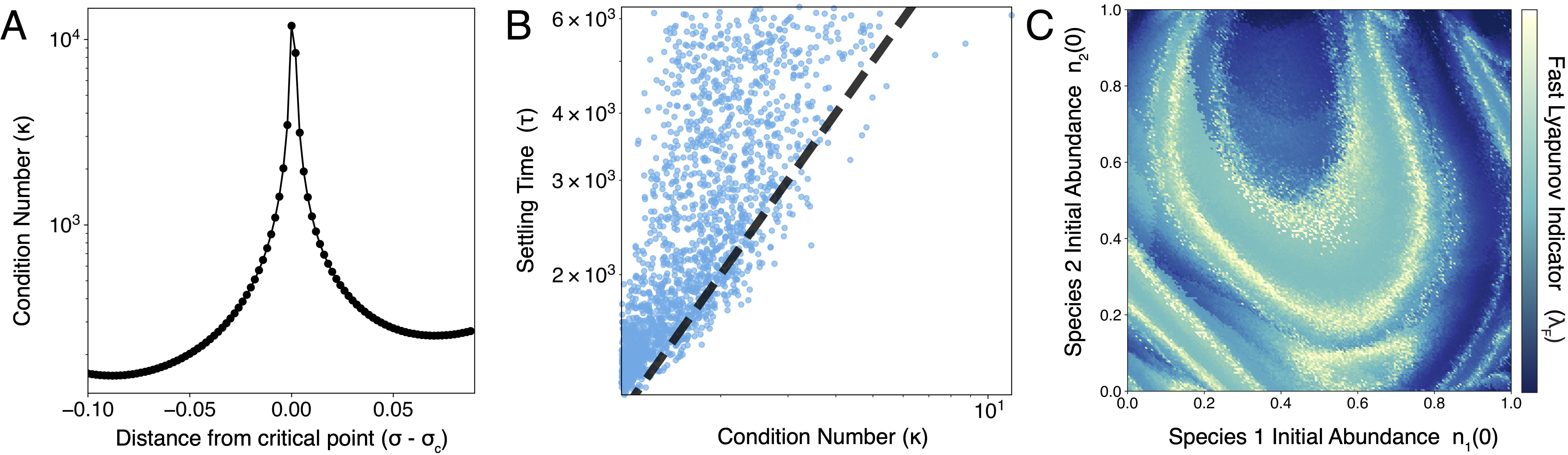}
\caption{
\textbf{Complex transients in an alternative ecosystem model.} 
(A) The condition number as an ecosystem model transitions from noninvasible global stability to multistability. 
Settling time $\tau$ versus condition number $\kappa$ for $10^{4}$ random communities.
(B) Caustics in the Fast Lyapunov Indicator ($\lambda_F$) versus initial conditions on a two-dimensional slice through the $N$-dimensional space of initial species densities. 
}
\label{altmodel}
}
\end{figure}

In our formulation of the generalized Lotka-Volterra model \eqref{lv_supp}, the role of the condition number $\kappa$ on the interaction matrix $A$ is explicitly encoded in the parameter $\epsilon \sim \kappa^{-1}$ via the structured interaction matrix \eqref{interact_supp}. Several recent studies instead parametrize $A$ in terms of a random ensemble of scaled interactions,
\begin{equation}
    A = -\mathbb{I} - \dfrac{\mu}{N}\boldsymbol{1} - \dfrac{\sigma}{\sqrt{N}} A'
\label{interact_alt}
\end{equation} 
where $\mathbb{I} \in \mathbb{R}^{N\times N}$ denotes the identity matrix, $\boldsymbol{1} \in \mathbb{R}^{N\times N}$ denotes a matrix of all ones, and the elements of $A' \in \mathbb{R}^{N\times N}$ are randomly sampled from a bivariate Gaussian distribution of the form
\[
    \langle A_{ij}' A_{k\ell}'\rangle = \delta_{ik}\delta_{j\ell} + \gamma \delta_{i\ell}\delta_{jk}
\]
so that $\langle A_{ij}'^{2}\rangle = 1$ and $\langle A_{ij}' A_{ji}'\rangle = \gamma$.

Analytical and numerical studies show that \eqref{interact_alt} exhibits a threshold $\sigma_c \equiv \sqrt{2}/(1 + \gamma)$ at which the system transitions from exhibiting multiple stable equilibria to a single, globally stable fixed point \cite{rieger1989solvable,bunin2017ecological,ros2023generalized}. We apply our analysis from the main text to this system. We simulate a random ensemble of ecosystems and find that, near the transition point between global stability and multistability, the condition number $\kappa(A)$ rapidly increases and produces long-transients associated with ill-conditioned dynamics  (Fig. \ref{altmodel}A). In this intermediate regime, the system exhibits the same phenomena we report in our main results, including scaling of equilibration (i.e. solving) time with the condition number  (Fig. \ref{altmodel}B), and the appearance of caustics and transient chaos associated with differing routes to equilibrium (Fig. \ref{altmodel}C).

\section{Effect of network connectivity on dynamics}
\label{app:sparse}

We repeat our scaling experiments for different values of the connectance $\rho$, corresponding to the total number of connections among nodes $C$ divided by the total possible connections $N(N-1)/2$. In Figure \ref{fig:connectivity2}, we find that the observed scaling of transient lifetime with condition number does not vary with connectance over two orders of magnitude. In Figure \ref{fig:connectivity}, we find that the pseudobasins associated with transient dynamics change in structure, but not intensity, as the connectance varies.

\begin{figure}
{
\centering
\includegraphics[width=0.5\linewidth]{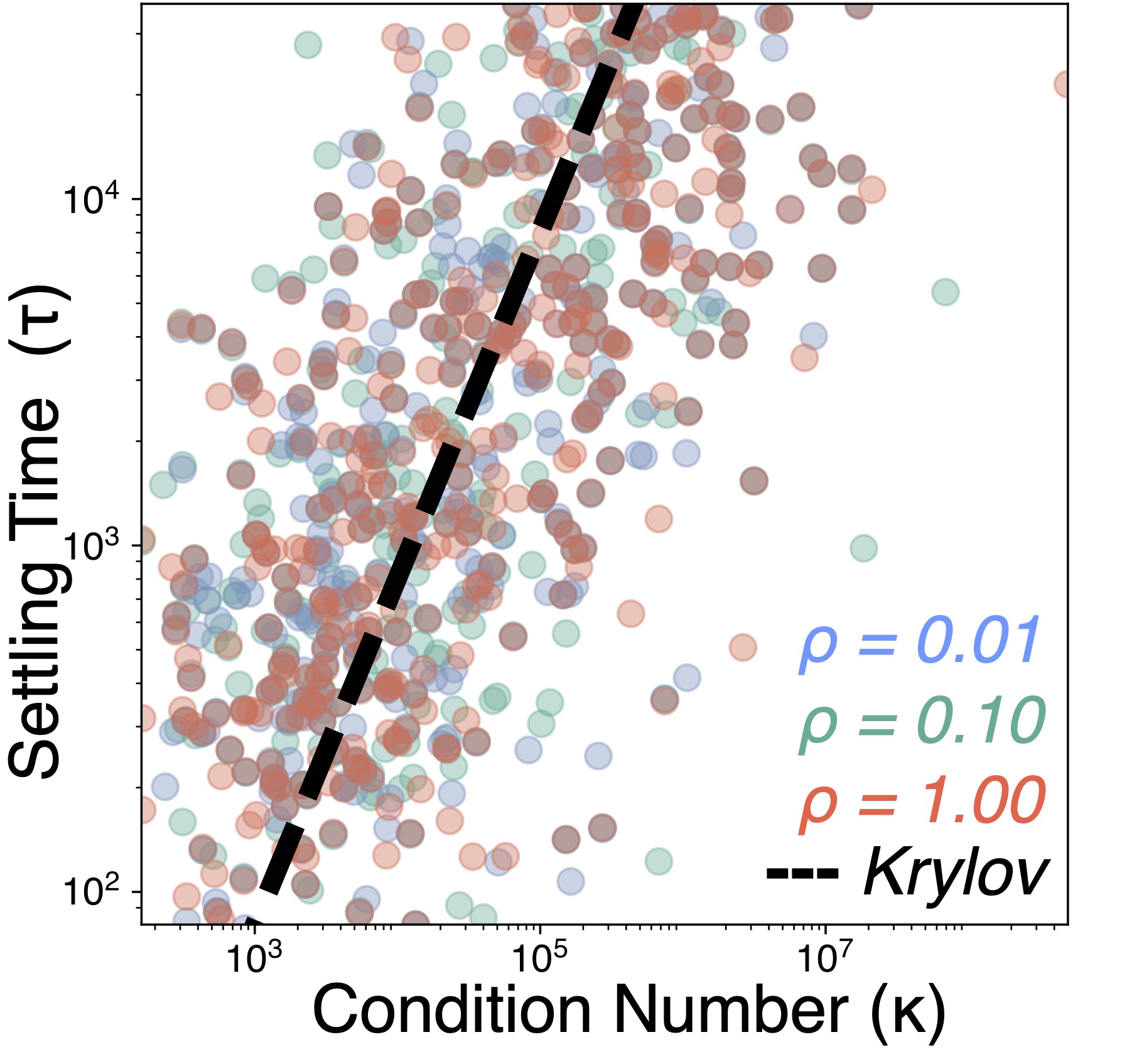}
\caption{
\textbf{The effect of network connectivity on scaling of transient dynamics.} 
Settling time $\tau$ versus condition number $\kappa$ for $3 \times 10^{3}$ random communities; dashed line shows the scaling expected for an iterative linear program solver. Colors indicate communities with different values of the connectance ($\rho$).
}
\label{fig:connectivity2}
}
\end{figure}

\begin{figure}
{
\centering
\includegraphics[width=\linewidth]{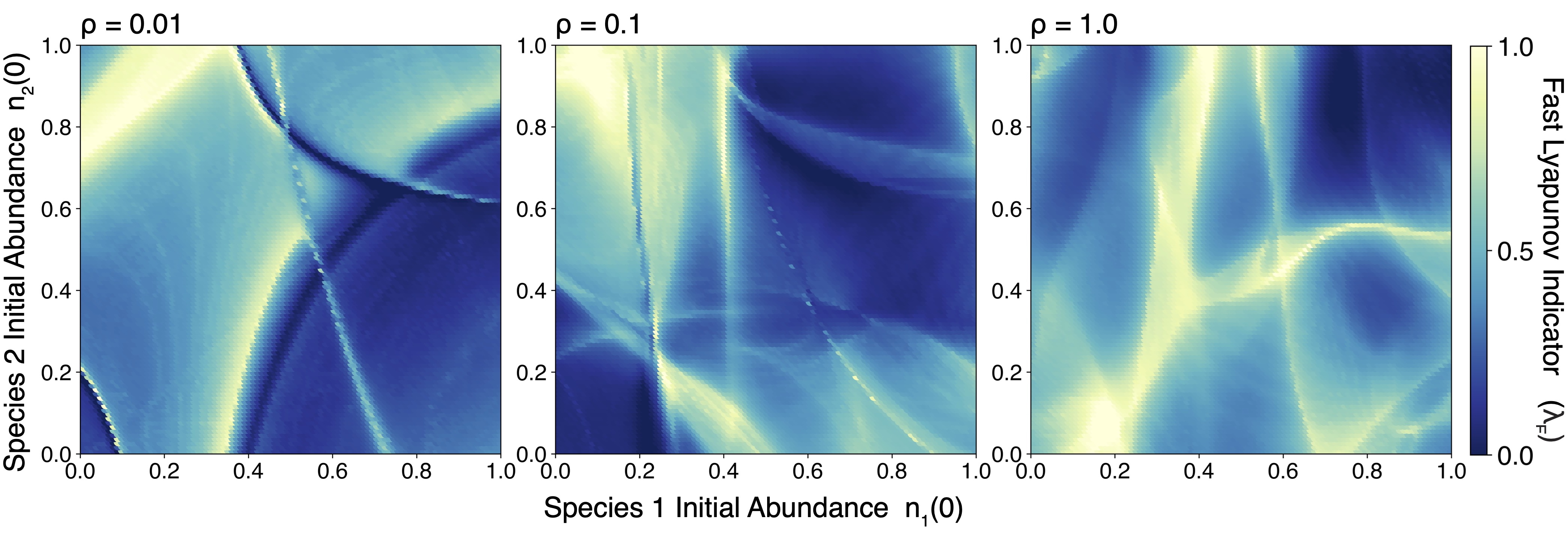}
\caption{
\textbf{The effect of network connectivity on transient dynamics.} 
Caustics in the Fast Lyapunov Indicator ($\lambda_F$) versus initial conditions on a two-dimensional slice through the $N$-dimensional space of initial species densities. Panels correspond to three levels of network connectance ($\rho$).
}
\label{fig:connectivity}
}
\end{figure}

\section{Effect of slow manifold dimension on dynamics}
\label{app:sparse}

We repeat our experiments by varying the relative dimensionality of the slow manifold, corresponding to $M/N$. 
In Figure \ref{fig:rank_scaling}, we find that the observed scaling of transient lifetime with condition number does not vary with connectance over two orders of magnitude. However, in Figure \ref{fig:rank_basins}, we find that the pseudobasins associated with transient dynamics change in structure and intensity. In general, as the slow manifold becomes a larger dimensional slice through phase space, more routes trap trajectories along the slow manifold, leading to sharper caustics in the fast Lyapunov indicator $\lambda_F$.

\begin{figure}
{
\centering
\includegraphics[width=0.5\linewidth]{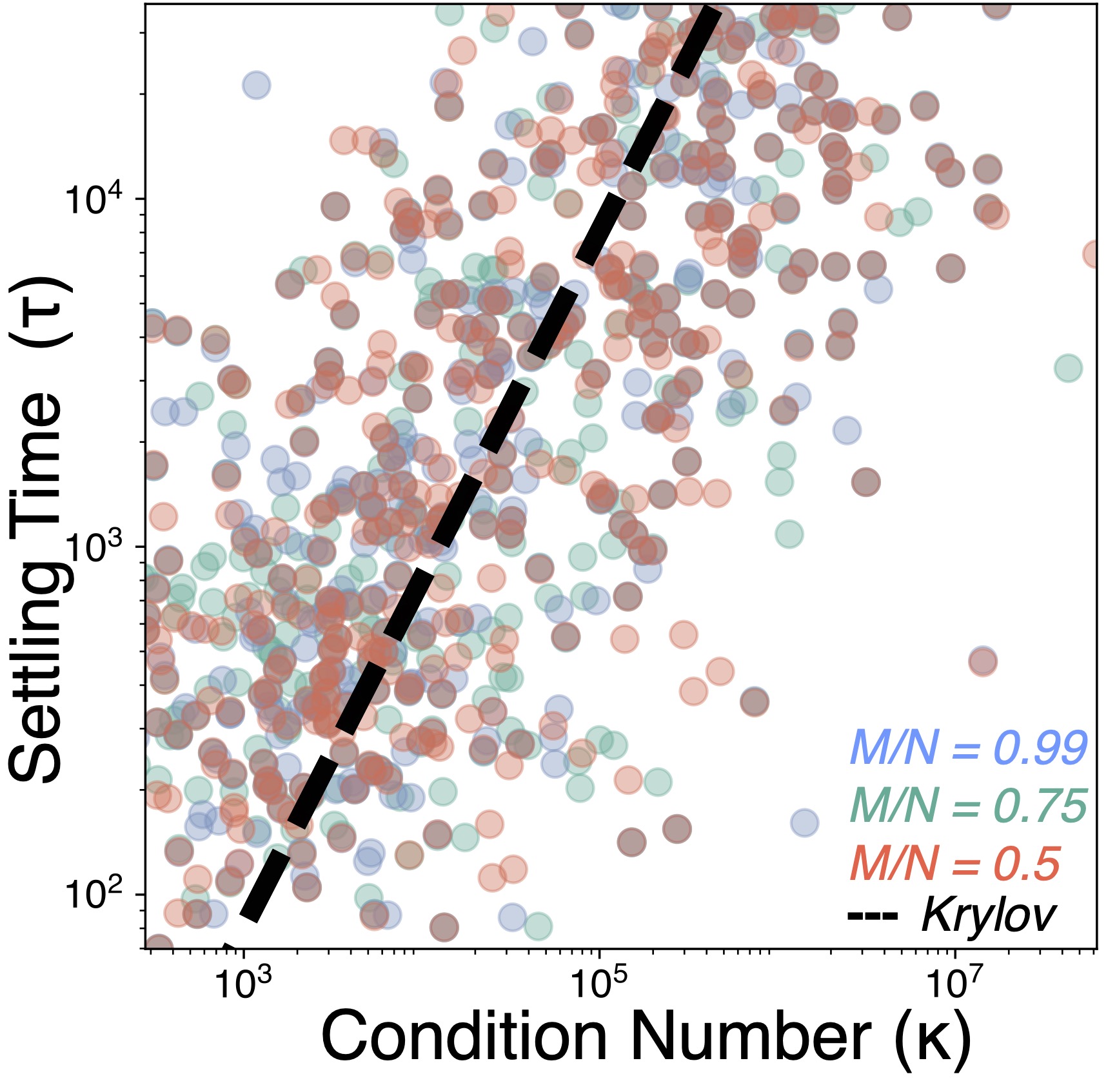}
\caption{
\textbf{The effect of slow manifold dimension on scaling of transient dynamics.} 
Settling time $\tau$ versus condition number $\kappa$ for $3 \times 10^{3}$ random communities; dashed line shows the scaling expected for an iterative linear program solver. Colors indicate communities with different slow manifold dimensionalities ($M/N$).
}
\label{fig:rank_scaling}
}
\end{figure}

\begin{figure}
{
\centering
\includegraphics[width=\linewidth]{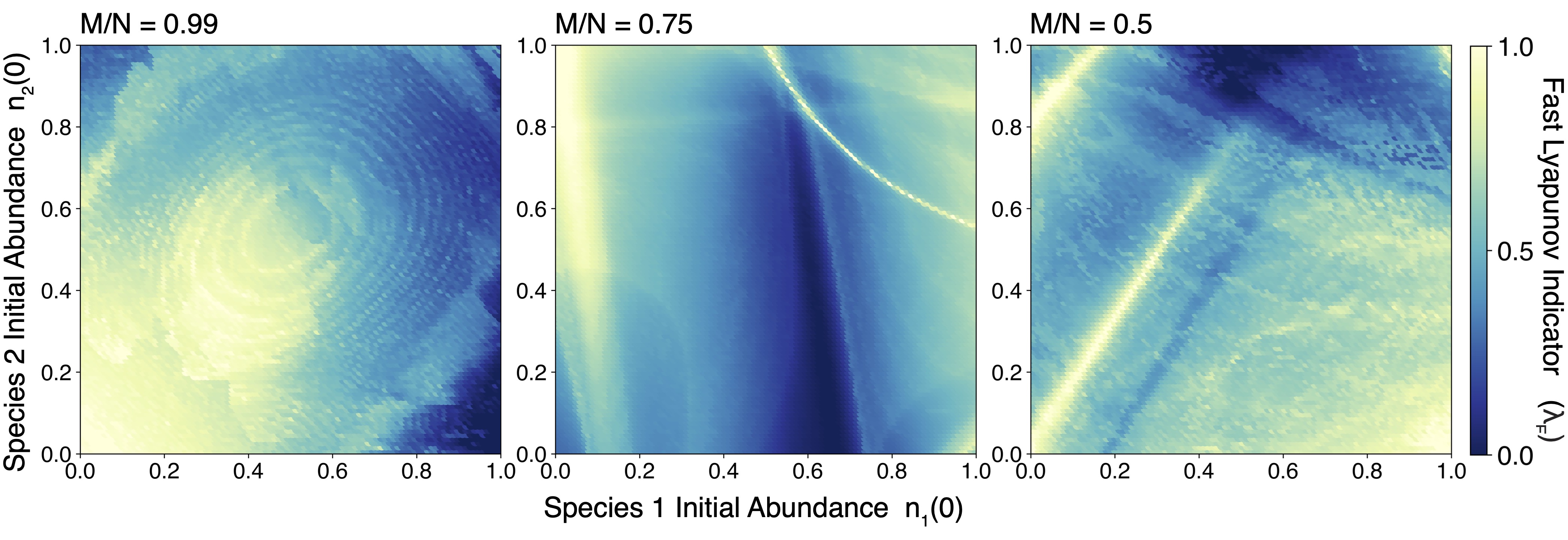}
\caption{
\textbf{The effect of  slow manifold dimension on transient dynamics.} 
Caustics in the Fast Lyapunov Indicator ($\lambda_F$) versus initial conditions on a two-dimensional slice through the $N$-dimensional space of initial species densities. Panels correspond to three values of the slow manifold dimensiona ($M/N$).
}
\label{fig:rank_basins}
}
\end{figure}

\end{widetext}

\end{document}